# What is a leader of opinion formation in bounded confidence models?

E. Kurmyshev [a, b *] and H. A. Juárez [a]

[a] Departamento de Ciencias Exactas y Tecnología, Centro Universitario de los Lagos, Universidad de Guadalajara. Lagos de Moreno, México
[b] Centro de Investigaciones en Óptica A. C. León, México.
*Correponding author

E-mail addresses: ekurmyshev@culagos.udg.mx, kev@cio.mx (E. Kurmyshev), hjuarez@culagos.udg.mx (H.A. Juárez)

**Abstract**

Taking a decision in democratic social groups (societies) is based on the opinion of the majority or on the consensus, if it is the case. Personalities and opinions of individuals constituting a social group in different stages of its development are used to be heterogeneous. The difference in opinions is indeed a motive to changes in a social group. So, the study of opinion dynamics is of great interest in analyzing social phenomena. Among the different models of opinion dynamics, bounded confidence models have been studied in different contexts and shown interesting dynamics, particularly in clustering, polarization and fragmentation of opinions, and influence of extremists [1 - 3]. In [4] we proposed a new bounded confidence model and studied the self-formation of opinion in heterogeneous societies composed by agents of two psychological types, concord (C-) and partial antagonism (PA-) agents. In this work we study the influence of "leaders" on the clustering of opinions in small world (SW) networks, starting the opinion dynamics from the uniform initial distribution of opinions in the network. Mixed C/PA-societies along with the pure C- and PA-society are studied. The influence of the leader's connectivity in the network, his toughness or tolerance and his opinion on the opinion dynamics is studied as a function of the initial opinion uncertainty (tolerance) of the population. Numerical results obtained with leaders at low, high and average tolerance show complex bifurcation patterns of the group opinion; a decrease or even the total lost of control of the leader over the society is observed in different intervals of tolerance of agents in the case of C/PA-societies. We found that in the C-society a leader showing high opinion tolerance has more control over the population. In the PA-society a leader changes the bifurcation pattern of group opinion in a drastic and unexpected way, contrary to the common sense, and generates stronger polarization in the opposite opinion groups; the connectivity of the leader is an important factor that usually improves the adhesion of agents to the leader's opinion. A low tolerance (authoritarian) leader has greater control over a PA-society than that of a high tolerance (democratic) one; the opposite result is obtained in the C-society.

## 1. Introduction

One of the major efforts in the field of complex networks is directed to the studying of social phenomena on networks, where the individuals (agents) are the nodes of a network and the relations between them are the links. Most of the social phenomena can be understood or interpreted through the dynamics of opinion formation. Among them the search for agreements and the reaching of consensus are important aspects of social group dynamics that finally enhance its impact on the society. A social group is used to have a leader (or leaders) which influence the "social" life, in particular through the opinion formation. In this work we model and study a leader's influence over the opinion formation in heterogeneous social groups.

The leadership can be defined in a variety of manners, but most of the definitions consider the leadership as the ability of a person to convince the others to reach certain objectives [5, 6]. In order to convince persons a leader executes his power using the following "tools" basically [6-9]. Compensation is done when a leader disposes and provides psychological and/or economical stimulus in exchange for the loyalty to the idea. Coercive power is based on fear, and is executed through brutal force, threats and punishments. Legal power is associated with the role of a chief by his formal position. Natural or charismatic power is based on the admiration of a leader by the followers. Finally, a leader can be an expert in a particular area or topic, which has a mayor capacity than others to analyze, control and execute the tasks.

Each leader has its personal style of guiding a group. In general, the style depends on the personality of the leader, even though this can be modified by personalities of the subordinates or followers, the nature of the task and/or the structure of a company. A leader possesses a personal social-psychological style of executing a power; it can be authoritarian, paternalistic, advisory or democratic [6-9]. In addition, sociologists used to distinguish among leaders of social groups or organizations, leaders of social networks and mass media, even though there is much in common among those. Any model of social behavior and/or opinion formation is limited, so it is hardly possible to bring together all features mentioned above and study the leaders' influence in a mathematical model. How and which of the above mentioned characteristics of a leader can be incorporated in a model of opinion dynamics? Under what conditions can an agent, possessing some leader's qualities in terms of the variables and parameters of the model, act as a leader?

Among the works on the social groups and networks there is a relatively small number of works on the mathematical modeling of a leader influence on the opinion formation. The social-psychological aspect of a leadership is used to be studied in social sciences and those are descriptive mainly. They propose or define the attributes of a leader and describe his/her behavior in the society [6-12]. The authors of [10] outline a method to accelerate the dissemination of innovations using opinion leaders, present optimal procedure and report on computer simulations that show how much faster diffusion occurs when initiated by opinion leaders. The topic of diffusion of opinions in networks is also discussed in [11]. Several attributes related to opinion leaders are reported in [12] after the revision of the concept of opinion leadership. On the other hand, the mathematical modeling of opinion formation is based mainly on the social physics models that take into account dynamical nature of social interactions and different roles of individuals [2, 13-15]. Those are physical models adapted for the social phenomena, in particular, the opinion formation. Being limited for the description of opinion formation, they have reduced capacity for incorporating leaders of different types in the model [16-22]. Another type of mathematical models are needed, they should arise from the social and psychological behavior of individuals. Mathematical variables and parameters of the model should describe the concepts and notions of social science with high fidelity from the very beginning of the modeling. Combining socio-psychological approaches with the quantitative mathematical tools and statistical physics can help to achieve reliable insights. Both approaches are important to improve the predictive capability of the opinion formation models. To our opinion, the bounded confidence, relative agreement models (BC/RA) of opinion formation [1-4] are close to the last requirement.

Self-organization of opinion not influenced by leaders was studied extensively in the frame of C/PA model [4], giving interesting bifurcation patterns in opinion space as a function of opinion tolerance of agents. This agent-based model allows simulating the opinion formation in a mixed society constituted of agents of the two psychological types, concord and partial antagonism agents. Models of opinion dynamics are usually composed of the following basic elements: a) Opinion Space, either discrete or continuous; b) Updating Rule describes the change of opinion of a given agent in dependence on opinions of agents related to him; c) Updating Dynamics characterizes the interaction between agents, that can be either pair or group interaction and d) Social Network describes relationships between agents (social organization). Each and every of these elements contribute different way to the opinion formation in a society. Since the leadership is an intrinsic characteristic of a social group, the influence of a leader on the opinion evolution and formation is an important thing to study. In [4] was found that the opinion formation to be similar in the networks of the following three types: Erdös–Rényi random graphs, Watts-Strogatz small-world networks and complete graphs. In this work we limited our study to the small world networks only.

The leadership is the ability of an individual to convince the others to reach certain objectives. Leader of opinion has to influence the opinion of "subordinates-followers" in a way that they adopt the vision of the leader; he/she has to be firm. Because a leader maintains a course in order to approach the goal, we consider as a leader an agent which maintains an invariable opinion in the process of social opinion formation. Another distinguishing feature that helps a leader to execute the power is its relatively high connectivity into the social network. In addition, we control the leader's tolerance to simulate authoritarian, paternalistic, advisory or democratic leader. The social power of a leader is modeled partially via the connectivity of the leader in the network and through the relative agreement of the updating rule. In order to see separately different effects of the leader's influence on the opinion formation, in this work we consider a unique leader of the social group. It should be noted that a mass media can be treated somehow as a leader of opinion by itself, or as a tool of a leader.

This paper is organized as follows. In Section 1 we give a brief qualitative description of a leader and trends in modeling of opinion dynamics in social groups with leaders. A model of opinion dynamics in a mixed C/PA society is presented in Section 2. In Section 3 we incorporate a leader into the C/PA model of opinion dynamics, using variables and parameters of the model. Section 4 presents results of computer simulation. Finally, conclusions and discussion is given in Section 5.

## 2. C/PA model of opinion dynamics

Social groups are constituted of persons of different psychological types [14, 15]. In order to capture this important feature of social organization, in [4] was proposed a model of opinion in a mixed C/PA society which consists of agents of two psychological types, partial antagonism (PA-) and concord (C-) agents. The model uses interval $S = [-1, 1]$ as a continuous opinion space. In addition, each agent $i$ is characterized by his opinion uncertainty or tolerance $u_i \in (0, 1]$. An opinion segment $s_i = [x_i - u_i, x_i + u_i]$ is assigned to each agent $i$. For two agents, $i$ and $j$, their opinion segments overlap if and only if $h_{ij} = \min(x_i + u_i, x_j + u_j) - \max(x_i - u_i, x_j - u_j) > 0$, where $h_{ij}$ is called an opinion overlap. In the C/PA-model the opinion and uncertainty of passive (receptive) agent $j$ are changed when $0 \leq h_{ij}$. Agents of the two types differs each other in the way they update their opinion in pair interaction. The updating rule for the opinion and uncertainty of a passive C-agent $j$ is as follows:

$$x_j := x_j + \mu_1 \left( \frac{h_{ij}}{2u_i} \right) (x_i - x_j) \tag{1}$$

$$u_j := u_j + \mu_2 \left( \frac{h_{ij}}{2u_i} \right) (u_i - u_j) \tag{2}$$

where the active agent can be either a C- or PA-agent. Note that the interaction of C-agents is always attractive similar to that of [3], that gives the name to concord agents. Agents with partial antagonism represent the other psychological type. Since real life interaction between persons is usually repulsive-attractive, we define the updating factor of this type of agents in such a way that opinions of two interacting agents can diverge. The new updating rule for the opinion and uncertainty of a passive PA-agent $j$ is as follows:

$$x_j := x_j + \mu_1 \frac{h_{ij}}{2u_i} \left( \frac{h_{ij}}{u_i} - 1 \right) (x_i - x_j) \tag{3}$$

$$u_j := u_j + \mu_2 \frac{h_{ij}}{2u_i} \left( \frac{h_{ij}}{u_i} - 1 \right) (u_i - u_j) \tag{4}$$

where active agent $i$ can be C- or PA-type. Note that the relative agreement $(h_{ij}/2u_i)(h_{ij}/u_i - 1) \in [-0.125, 1]$, thus the opinions of PA-agents can diverge from that of active agents when $(h_{ij}/u_i - 1) \leq 0$. The C/PA society of $N$ agents consists of two sub-populations, $M_C$ and $M_{PA}$ of $pN$ and $(1 - p)N$ size respectively, where a fraction of C-agents, $0 \leq p \leq 1$ is a parameter of the model. When an active agent $i$ interacts with a passive one $j$, agent $j$ changes its opinion and uncertainty following the updating rules for the C-agents, Eqs. (1) and (2), if $j \in M_C$ or those for the PA-agents, Eqs. (3) and (4), if $j \in M_{PA}$. Varying the value of $p$ from 0 to 1, we model mixed C/PA-societies with populations ranging through the pure C- to the PA-society.

The behavioral patterns of the two types (C- and PA-) proposed and exploited in our work are typical for social groups but each of them has no unique socio-psychological mechanism; other patterns can vary around in details and differ in mathematical representation depending on the variables and parameters of the model. The every day experience suggests that the simplest way to see psychological reaction of an individual is to refuse or accept the opinion of the "opponent". When the opinion of the "opponent" is accepted, the most probable result is that the opinions getting closer. Nevertheless, in order to accept an opinion of the other, an individual have to have an opinion close enough to that of the opponent and, in addition, to be confident in the opponent psychologically. The latter is a psychological factor that, to certain extent, should to be measured independently of the opinion. If an individual has no sufficient confidence in the "opponent", he will most probably refuse the opponent's opinion; if the unidirectional confidence is sufficiently high, the opinions most probably get closer. This is a tiny moment not simple to express mathematically. Being limited in the variables, we were forced to express this mechanism mathematically in terms of opinion and uncertainty (opinion tolerance) only, as it is described in Eqs. (3) and (4). It is difficult to encounter the situation when individuals having very close opinions and confidence each in other will reject the opinion of opponent. Because the refuse of opponent's opinion can hardly be total, we consider the situation of partial antagonism. In our model we treat completely confident individuals as a special group of concord agents. No socio-psychological study has been done to justify the specific mathematical relation (PA) used in these works. These are mostly heuristic relations.

### 3. Model leader in the mixed C/PA-society

We consider leadership as the ability of a person to guide individuals (agents) to a goal. So, *a leader needs to be able to communicate ideas to agents and convince them* [5, 6]. To communicate implies that the leader can interact with a relatively large portion of population, directly or implicitly. To convince implies having some kind of "power" of opinion manipulation. This "power" can be either natural (charismatic leaders), given by some degree of reputation or experience, or implemented by some means such as a reward or even coercion. Another important characteristic of the leader is his ability to take into consideration other's opinion. From this point of view, a leader can be democratic or autocratic (dictator). Since any opinion model is limited, we represent these features in the following way. Our leader is an agent connected with a significant part of population; his degree of connectivity with other agents of the society, denoted by $d_l$ runs through 0.2 to 1. Because a leader has a goal (an objective) and strong conviction to reach it, we assign an invariable opinion to the leader, denoted hereafter by $x_l$. This can be viewed as a straightforward strategy of the leader. The "democracy level" of the leader is represented by his uncertainty or tolerance, $u_l$. We suppose that a dictator (authoritarian leader) has invariable low tolerance $u_l = 0.2$, whereas a democratic leader has invariable, relatively high tolerance $u_l = 0.5$. We also expect that the leader's tolerance can be equal to that of the rest of population, $u_l = U$, where $U$ is the average of initial tolerance of population; in that case the leader is called populist.

Social power is rather complicated socio-psychological notion. In this model it is regulated by the two magnitudes. First, it is the degree of connectivity of a leader in the network that is a reflection of his communicability and his social power. Second, the relative agreement between agents "$i$" and "$j$" regulates the magnitude of changing in the opinion of "passive" agent "$j$". Therefore, it represents the social power of an active agent "$i$", and, in particular, of a leader that is considered as an active agent always. For the passive agent "$j$" of the C-type, the relative agreement is equal to $a_C = h_{ij} / u_i$, Eq. (1). When the passive agent "$j$" is of the PA-type, the relative agreement is equal to $a_{PA} = (h_{lj} / u_l - 1) h_{lj} / 2 u_l = (a_C - 1) a_C / 2$, Eq. (3).

In practice, every leader usually uses a combination of powers, but not the only one. Different combinations of connectivity and tolerance can be considered as a combined, integral representation of leader´s powers. For example, in a professional group, probably the most important of the leader´s powers is the professional reputation, but charisma and reward can not be excluded and certainly are present; they enhance the reputation power. In this model, the reputation can be considered as a combination of a relatively high tolerance and connectivity. Charisma is a psychological quality, and in the frame of this model it can be partially attributed to a relatively high tolerance; the latter is a feature of a populist also. As for the connectivity, it can be considered as a part of a strategy that can be used by a leader to reach a desired result.

When a leader is a "dictator" (coercive power, perhaps enhanced with a reputation and rewards), he used to have a low tolerance (no charisma) and fixed opinion; and, most probably, he will need high connectivity and/or hierarchical structure of the society to control the opinion or, at least, the opinion expression.

Note that the way we conceive and implement the leader can be used to model different situations where a kind of leadership is necessary, for example, leaders of social groups or organizations or leaders of social networks. In particular, adjusting parameters attributed to the leader we can simulate the effect of mass media in the opinion dynamics of population.

## 4. Results of computer simulation

In this work we are mainly interested in the asymptotic, stationary distribution of opinion in the society. This implies that a leader has quite enough time to work with the agents and convince them if it is the case. Since we are interested in the final state of the opinion, we do not study opinion convergence in time, leaving this important detail for future work.

In order to appreciate changes in the opinion formation caused by the presence of a leader in a social network, we studied numerically the patterns of opinion distribution on the Small World (SW) networks of PA- and C-agents, in function of the parameter $U$ ranging through 0.3 to 1.2, where $U$ is the average of initial tolerance of population. Results for $U \geq 1$ has theoretical but not practical sense, because it is hardly expect that almost all population has so big opinion uncertainty in the opinion space $x_i \in [-1,1]$. Leader's attributes, his connectivity $d_l$ to the network, tolerance (uncertainty of leader's opinion) $u_l$ and opinion $x_l$ were varied in a wide range of values. Note that the opinion $x_l$ is kept invariable during the evolution of the society opinion.

We wonder to know, how does a model leader change the bifurcation pattern of opinion compared to that in the society without leader? In order to simulate the influence of a leader on the opinion formation in the mixed C/PA-model we use the following methodology. The society of individuals is a small world network with $N = 1000$ agents. The state variables of the society are the opinion $x_i$ and the uncertainty $u_i$ of individuals (agents), $i = 1,2, …N$. Uniform probability distribution is used to choose the initial opinion of agents from the interval [-1, 1] and uncertainty from the interval $[U − 0.2, U + 0.2]$, where $U$ is the average initial opinion uncertainty of the population. We add to the network a leader with the opinion $x_l$ and uncertainty $u_l$ by connecting him randomly to $d_l N$ agents, where $d_l$ is the connectivity of the leader in the network. At each time step, we choose randomly $N$ edges (pairs of coupled agents) of the network. If one of these agents is the leader, then his partner is chosen to be the passive agent always; this guarantees that the leader does not change his opinion and uncertainty. If neither of the pair of agents is a leader, then one of them is selected at random to be the active agent $i$, while the other is considered to be the passive one, $j$. In our study we suppose that, in the process of opinion evolution in time, the dictator has invariable low tolerance $u_l = 0.2$, whereas a democratic leader has invariable, relatively high tolerance $u_l = 0.5$ through all the range of the average initial opinion uncertainty $U$. As for the populist leader we expect that his/her tolerance is similar to the rest of population, $u_l = U$.

The value of the opinion overlap $h_{ij}$ is computed, and we update variables of the passive agent if and only if $h_{ij} \geq 0$. As mentioned above, the composition of heterogeneous society is regulated with the parameter $0 \leq p \leq 1$. When an active agent $i$ interacts with a passive one $j$, agent $j$ changes its attitude according to the C-type rules if $j \in M_C$, or the PA-type rules if $j \in M_{PA}$. The value of the parameter $U$ runs through 0.3 to 1.2 at the step of 0.01. At each value of $U$, we execute 350 iterations (time steps). Because we are basically interested in the steady state of opinion dynamics, we plot the density of asymptotic opinion distribution as a function of parameter $U$. The results shown in the following figures represent an average over 50 simulations for each value of $U$.

Figure 1 shows bifurcation patterns for the societies without leaders as the reference; those are calculated just under the same conditions as for the societies with leaders, presented in this work. This comparison is instructive for the studying of natural competition between the self-organization of the society and the influence of a leader on the opinion formation.

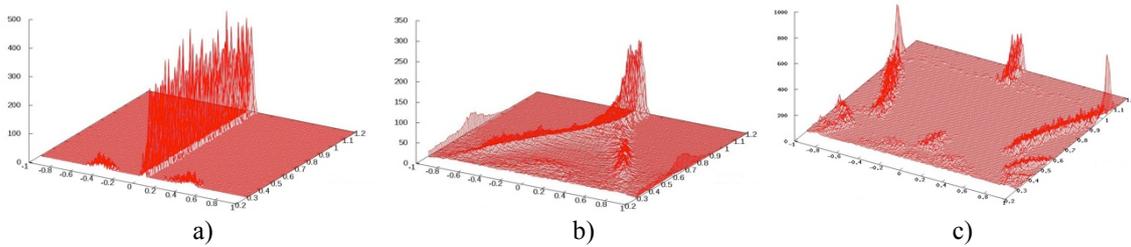

Figure 1. Bifurcation patterns of opinion in the C/PA-societies without leaders on the SW networks: a) C-society ($p = 1$); b) Mixed C/PA-society ($p = 0.5$) and c) PA-society ($p = 0.0$)

Our experiments pretend to answer the question: Under what conditions does an agent act as a leader of opinion? That is, when he attracts opinions of the majority of individuals? To answer the question, in computer simulations we varied the following parameters of the model. For each C/PA composition of the society running through 0/100, 20/80, 40/60, 60/40, 80/20 and 100/0 ($p = 0, 0.2, 0.4, 0.6, 0.8$ and 1), the leader's parameters were given the following values: opinion $x_l = 0, 0.2, 0.4, 0.6$ and 0.8; his tolerance $u_l = 0.2, 0.5$ and $U$; his degree $d_l = 0.2, 0.4, 0.6, 0.8$ and 1. Bifurcation patterns and plots of temporal evolution of opinion demonstrated a variety of interesting details and tendencies; those are hardly possible to comment in detail in one article. So, in order to resume most of the results in this article we selected only few of them, those being most representative cases in our opinion.

Behavior of concord agents is rather simple and predictable. In Figure 1 we see that the consensus centered at the opinion $x = 0$ prevails in the C-society without leader, for the uniform initial distribution of opinion at values $U \geq 0.4$, although the small fragmentation of opinion is observed at values $U$ less than 0.4. As expected, we found that the centrist leader, $x_l = 0$, of any type only enhances the convergence to the unique opinion $x = 0$. The consensus is always observed in the C-society, for the democratic ($u_l = 0.5$) and the populist ($u_l = U$) leader; two small groups of separatists with opinions $x =$

0.5 and $x = -05$ are observed at $U$ close to 0.3 with a dictator ($u_l = 0.2$) at all values of his degree. So, the dominant opinion distribution in the C-society is the consensus, $x = 0$ in the presence of a centrist leader of any type, having the degree $d_l = 0.2, 0.4, 0.6, 0.8$ and 1; for that reason we omit the corresponding plots.

The patterns of asymptotic opinion distribution in the PA-society strongly differ to that of the C-society, with and without leaders (for example, compare the bifurcation patterns of Figure 1a and 1c). Computer simulations on the opinion formation in the PA-society with the centrist leader show that the asymptotic opinion distribution over the network is similar, except small quantitative details, in the society guided by the democratic leader, $u_l = 0.5$ and by the populist, $u_l = U$. For that reason we present opinion distribution only for the democratic leader, see Figure 2, line A. In both cases we observe strong generation of opposed extremists and incapacity of a "leader" to dominate the opinion of self-organized regular agents, in the interval $0.5 < U < 1.0$ (see Figure 1c also). These effects are modified quantitatively (but not qualitatively) with the degree of a leader (see Figure 2, line A). Nevertheless, a low tolerance leader (dictator) used to dominate the society opinion at a relatively high connectivity, $d_l > 0.6$; that is not the case at $d_l = 0.2$ when the dictator has little control in the interval $0.3 < U < 0.6$ and looses it in the interval of $0.6 < U < 1.0$ (see Figure 2, line B).

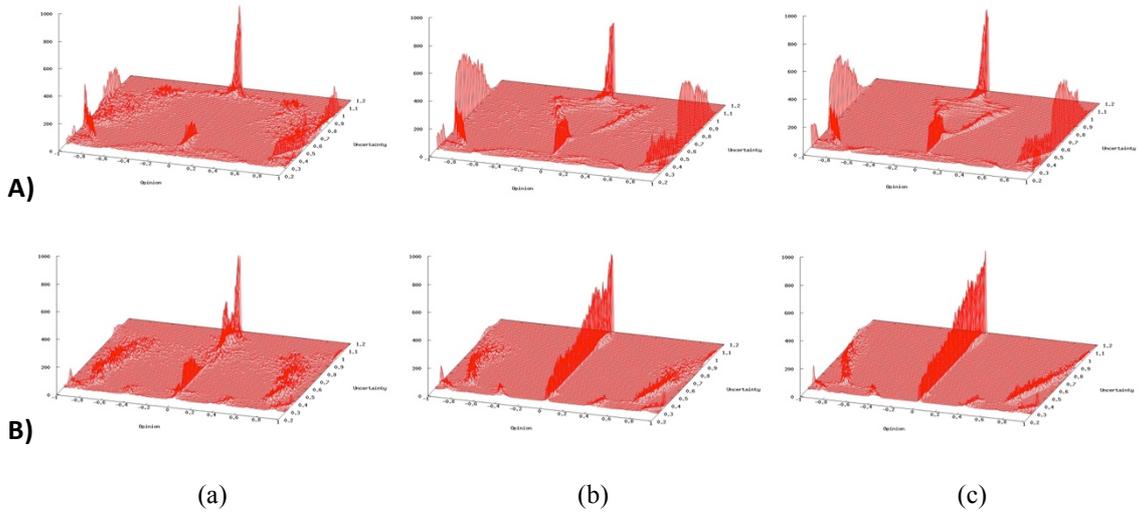

Figure 2. Bifurcation patterns of opinion in the PA-society, on the SW networks, guided by a centrist democratic leader, $u_l = 0.5$, line A and a "dictator", $u_l = 0.2$, line B: column (a) $d_l = 0.2$; (b) $d_l = 0.6$ and (c) $d_l = 1.0$

In real life it is hardly possible to meet homogeneous social groups composed of agents of the same psychological type. So, having in mind specific features of the opinion formation in pure C- and PA-society we wonder to see the influence of a leader on the opinion formation in mixed societies. Even though we have results of computer simulation for the fractions of C/PA agents through 0/100, 20/80, 40/60, 60/40, 80/20 and 100/0, we consider appropriate to show and analyze here the plots in the case of C/PA-40/60 that reveals general tendencies, Figure 3.

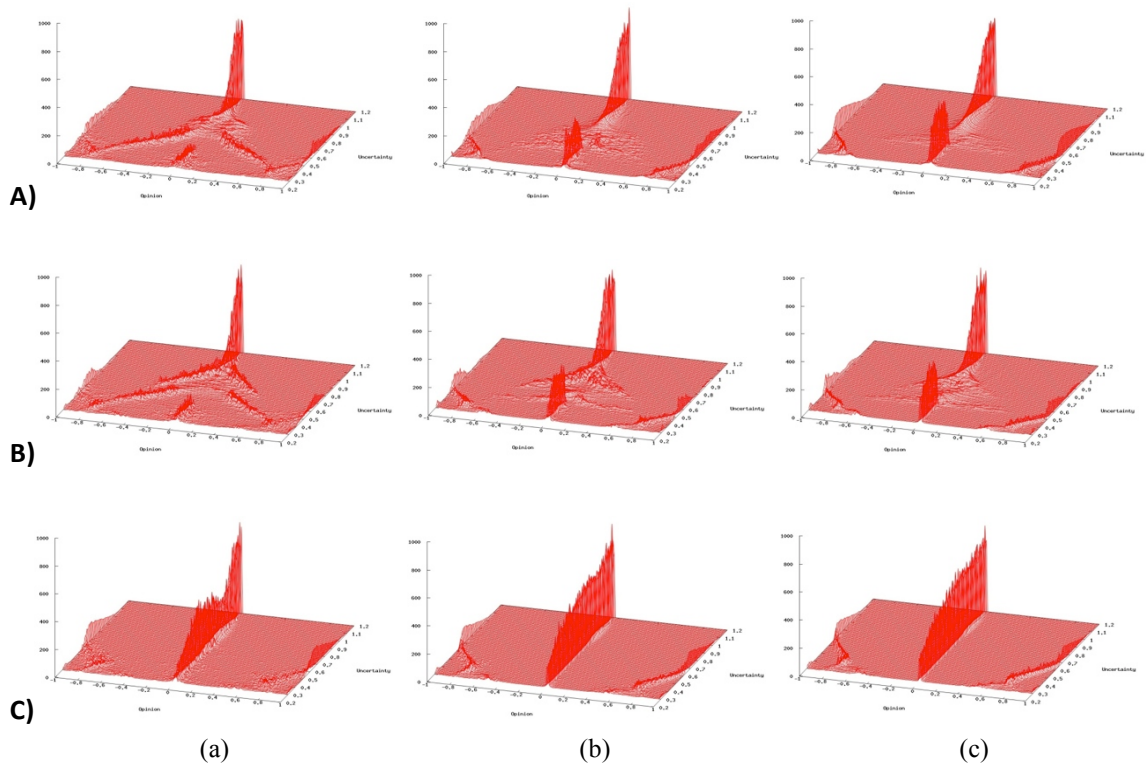

(a)                  (b)                  (c)

Figure 3. Bifurcation patterns of opinion in the C/PA-40/60 society, on the SW networks, guided by a centrist democratic leader, $u_l = 0.5$, line A; populist leader $u_l = U$, line B and a "dictator", $u_l = 0.2$, line C: column (a) $d_l = 0.2$; (b) $d_l = 0.6$ and (c) $d_l = 1.0$

Asymptotic opinion distribution over the network is similar, except some quantitative details, in the society guided by a democratic leader, $u_l = 0.5$ and a populist $u_l = U$; see Figure 3, lines A and B. Nevertheless, the low tolerance leader ("dictator") shows surprisingly different results in opinion formation; those are given in Figure 3, line C. Dictator demonstrates much better performance; he dominates the society opinion for $U > 0.4$ and does not loose the governability in any interval of $U$ in contrast to that of the populist or democratic leader. Another interesting feature of this opinion dynamics is that centrist leaders of any tolerance used to generate extremists. Bifurcation patterns are modified in function of the leader's degree (connectivity). Those general tendencies are also observed in the mixed societies at C/PA ratio equal to 20/80, 40/60, 60/40 and 80/20, with notable variations in quantitative details through the society composition.

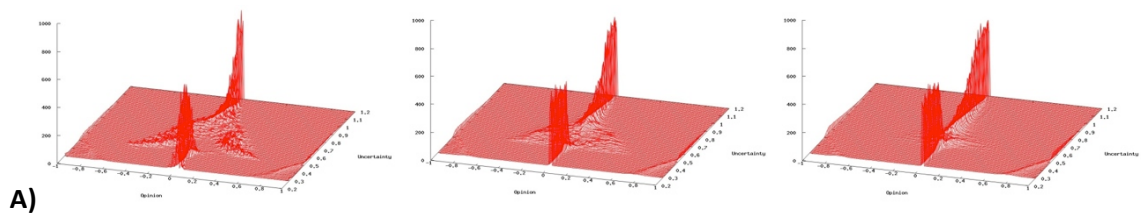

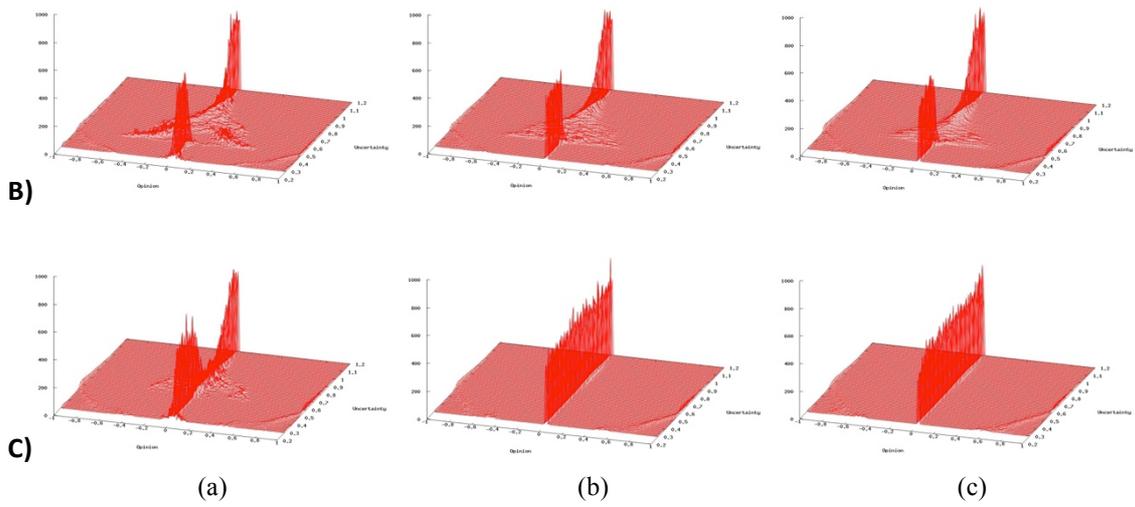

Figure 4. Bifurcation patterns of opinion in the C/PA-80/20 society guided by a centrist democratic leader, $u_l = 0.5$, line A; populist leader $u_l = U$, line B and a "dictator", $u_l = 0.2$, line C: column (a) $d_l = 0.2$; (b) $d_l = 0.6$ and (c) $d_l = 1.0$

From Figure 4 we see that the presence of even 20% of PA-agents changes significantly both the dynamics and the final distribution of opinion in the mixed C/PA-80/20 society compared to that of the pure C-society; in particular, we observe intervals of $U$ where the centrist "leader" notably looses control over the opinion of regular agents. It is a result of competition between the self-organization of opinion of agents and the open loop control of the leader over their opinions. There is also an initial, symmetric repulsion of agents from the leader and then their self-organization in the opposite groups. Note that Figure 4 has notable correlation with the case of C/PA – 40/60, Figure 3.

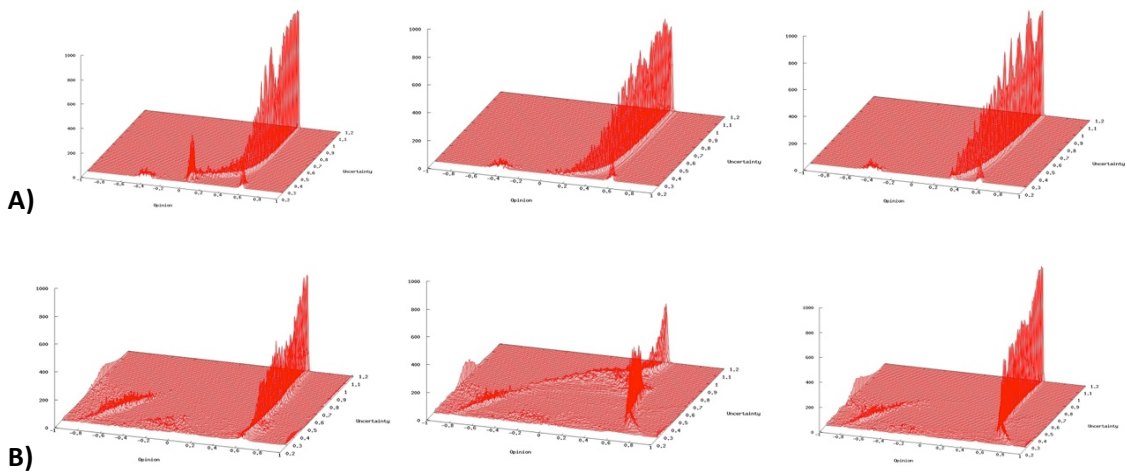

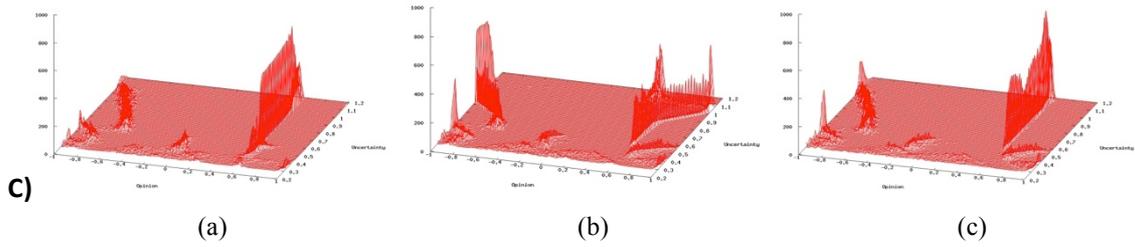

**C)**

(a)          (b)          (c)

**Figure 5.** Bifurcation patterns of opinion in the SW network as a function of $U$, for the leader's opinion $x_l = 0.6$ and the degree $d_l = 0.2$. Row A, pure C-society ($p = 1$); row B, mixed C/PA-society ($p = 0.4$) and row C, pure PA-society ($p = 0$). Columns: (a) dictator, $u_l = 0.2$; (b) populist, $u_l = U$ and (c) democratic leader, $u_l = 0.5$.

Figure 5 shows, in series of plots, the influence of a leader, having well defined "right" orientation, $x_l = 0.6$ and a relatively low degree of connectivity, $d_l = 0.2$ on the opinion formation in the C-, PA- and mixed C/PA-40/60-society; leaders of the three types are considered. Comparing figures in columns (a) and (c) in Figure 5, we see that in general a democratic leader and a dictator, with relatively low degree-connectivity (0.2), get similar results; those differ from that of the populist (compare columns (a) and (c) versus (b)). In the C-society, the populist, $u_l = U$ and democratic leader, $u_l = 0.5$ with relatively low degree (of connectivity), $d_l = 0.2$ have almost the same control over the final opinion of regular agents; they clearly dominate agents' opinions if the initial tolerance (uncertainty) of the population $U \geq 0.4$, the distribution of opinion having the bifurcation and fragmentation of opinion at values $U < 0.4$. The dictator, $u_l = 0.2$, also gets control over the opinion of agents, but in the societies with the higher tolerance, $U \geq 0.5$. For the less values of $U$, the group opinion suffers notable fragmentation and the dictator does not get control over it.

In the C/PA-40/60-society (Figure 5, line B), a dictator and a democratic leader get similar control over the agents' opinions, and they are quite efficient in that issue for the societies with agents whose tolerance $U \geq 0.5$. But the populist is less efficient and he gets only partial control over the opinion of population in some intervals of the initial tolerance of population, $U$. Leader of any type induces generation of the left extremists, $x = -1$, in the interval $0.6 \leq U \leq 1$.

In case of the PA-society, the patterns of opinion bifurcation look more complicated, Figure 5, line C. The control of agents' opinion by a dictator and a democratic leader is similar, except quantitative details; in general, they control efficiently the opinion of population at values $U \geq 0.5$. As for the populist, his control of opinion is little efficient. In the interval $0.6 \leq U \leq 1.2$, he gets a number of followers, but at the same time he provokes a generation of extremists, both left and right orientation (see Figure 5 (C, b)). Extremists opposing leader's opinion even exceed his followers in number. Bifurcation of opinion is observed at $U = 0.5$. As a result of this bifurcation a large group of left opponents, then becoming left extremists, emerges. In addition, a group of right oriented opponents, becoming right extremists, splits of the group of the leader's followers. This bifurcation pattern can be interpreted as the competition of the self-organization of opinions of agents and the open loop control of a leader over their opinions; a leader with opinion $x_l = 0.6$ at $U > 0.55$ attracts a supporting group of agents, which repulse the opinion of the opposite groups to the left and right extremes accompanied by the effects of self-organization of the extreme groups.

The plots of column (b) in Figure 5 show that the populist "leader" succeeds in dominating of public opinion in the C-society only; in mixed C/PA-societies there are intervals of $U$ where the populist looses the control over the public opinion, and this effect is more notorious in societies with a big PA-component of population. It should be noted that this tendency is observed for different values of $p$ and connectivity of the leader.

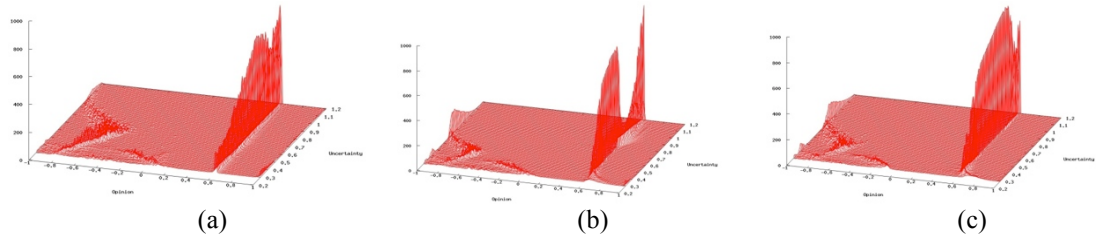

(a)          (b)          (c)

Figure 6. Bifurcation patterns of opinion distribution as a function of $U$ in the mixed C/PA-40/60 society for $x_l = 0.6$ and high degree of a leader $d_l = 0.8$: (a) dictator, $u_l = 0.2$; (b) populist, $u_l = U$ and (c) democratic leader, $u_l = 0.5$.

In order to show the influence a leader with a relatively high connectivity, in Figure 6 we present the bifurcation patterns of opinion in the mixed C/PA-40/60 society for $x_l = 0.6$ and $d_l = 0.8$ (high connectivity). Even though the dictator used to attract the opinion of a numerous group of agents for the society with the tolerance in the interval $0.3 \leq U \leq 1.2$, his control becomes efficient for $0.8 \leq U$ only (see Figure 6a). The populist is an efficient leader for the societies with a modest tolerance, $0.45 \leq U \leq 0.85$, but then he looses a control (see Figure 6b). From Figure 6c we see that the democratic leader joins the power of both, the populist and dictator, getting an efficient control over the society in the interval $0.3 \leq U \leq 1.2$. Democratic leader ($u_l = 0.5$) becomes a kind of a dictator for the agents with $U > 0.5$. For that reason his behavior is similar to that of the dictator for $U > 0.5$ (see Fig. 6a). In this situation the democratic leader controls opinions of agents; there is no gap, as it is observed in the case of the populist. Groups of left extremists and dissidents are observed in Figure 6. These tendencies where observed in different mixed societies with highly connected leaders. Comparing of Figure 6 with Figure 5, row B shows the importance of leader's degree for its performance. Note that leaders with high degree of connectivity in the networks can be interpreted as mass media to some extent.

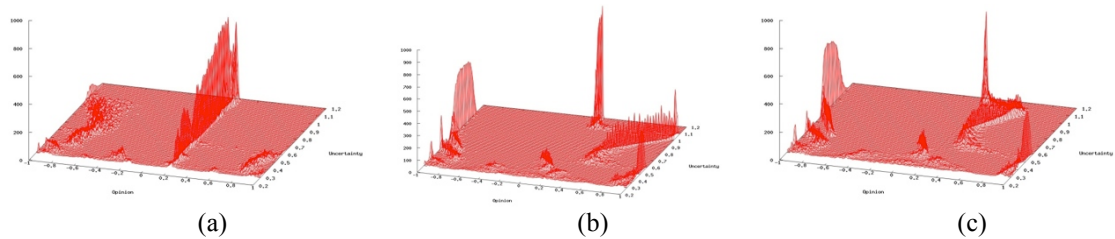

(a)          (b)          (c)

Figure 7. Bifurcation patterns of opinion distribution as a function of $U$ in a pure PA-society for $x_l = 0.2$ and $d_l = 0.2$: (a) dictator, $u_l = 0.2$; (b) populist, $u_l = U$ and (c) democratic leader, $u_l = 0.5$.

In order to see more details and analyze the influence of a leader with a modest right or left orientation, that is the intermediate case between the centrist and extreme oriented leader, we present in Figure 7 the series of graphics for the pure PA-society and leader's opinion $x_l = 0.2$ and $d_l = 0.2$; the latter means that the leader is almost centrist and has relatively low degree of connectivity. In this situation a dictator succeeds to get a control over the society, even though small groups of dissidents are observed for $U > 0.4$, Figure 7a. However, both the populist and democratic "leader" looses control over the society in the interval $0.3 \leq U \leq 1.1$, Figure 7b, c. Moreover, there are two groups of dissidents positioned in opposite sides of opinion space. The opinion distribution is not symmetric and the group of left dissidents and extremists is always larger than that of the right oriented agents (note that the leader is right oriented modestly). Bifurcation patterns of Figure 7 show that the characteristics attributed to the model leader are not always sufficient to guarantee a control over the society.

5. **Conclusions and discussion**

In this work we have proposed the model and studied the competition between the self-organization of opinion of agents in the network and the influence of a leader over the public opinion. We have introduced model leaders of the authoritarian (dictator), democratic and populist types in the heterogeneous C/PA model of opinion dynamics. A "leader" of the network opinion is an agent to which are assigned an invariable opinion and uncertainty (tolerance), and a relatively high degree of connectivity in the network. In this work we have shown, those are necessary but not sufficient conditions for an agent to be a leader. To be an efficient leader one needs to posses some more qualities than high degree of connectivity $d_l$ and invariable opinion $x_l$ and tolerance $u_l$; model leader, at least, has to assess the psychological type of participants (C/PA-agents) of the group and their tolerance $U$. In some respects the hubs could be thought as model leaders since they are high degree nodes in scale-free networks. Nevertheless, it is not intrinsic feature of an agent, but it is a topological property of the network that reflects a social position of the hub to some extent. In addition, hubs are numerous formations in the scale-free networks. The question is who is a leader? In this sense, any pair of hubs considered as leaders can have identical, similar or opposite opinions; these situations will produce different results. All of these qualities of hubs complicate, if not confuse the understanding of the leader's influence upon the opinion formation in a social group (network). In order to better the understanding of the mechanism of a leadership, in this work we have considered a unique leader in homogeneous networks. Small world networks were used to model the opinion dynamics of a society.

We have found that the opinion formation in a society is sensitive to both the psychological type of agents (C- and PA-agents) and the type of a "leader". In our experiments we observed the following tendencies.

Individual behavior of C-agents is rather simple, Eq. (1) and (2). Nevertheless, even in the pure C-society a control of public opinion by a leader is quite complicated. The opinion distribution in the C-society, in the presence of the dictator with the opinion in the interval $x_l \in [-0.4, 04]$, is the consensus centered at the leader's opinion $x_l$, for the leader's degrees of connectivity $d_l$ = 0.2, 0.4, 0.6, 0.8, 1 and $U \geq 0.35$. However, when a dictator has more extremist position, $x_l = \pm 0.6, \pm 0.8$, he almost looses control over the public opinion for $0.3 \leq U \leq 0.6$, getting a small number of followers at high degrees $d_l$ = 0.8, 1 only. For the societies with a high uncertainty (tolerance), $U > 0.6$, even a dictator with a radical opinion used to establish control and dominate public opinion. In the C-society the populist and democratic leader in general have better performance than that of a dictator in all range of parameters, the democratic leader being the best.

In general, model leader of any type has more problems with the control of opinion in the C/PA-societies with a notable portion of PA-agents. The population with at most 20% of PA-agents used to suffer multiple opinion fragmentations. There are intervals of opinion tolerance $U$, especially for societies with a notable PA-component, when the model leader looses control over the society (see Figures 2, 3, 4, 5, 7). In the case with the populist or democratic leader, we observe strong generation of opposed extremists and incapacity of a "leader" to convince regular agents, in the interval $0.5 < U < 1.0$ (see Figures 2A, 3A and 3B, 4A and 4B, 5B and 5C). In the societies with a notable portion of PA-agents the overall performance of the dictator in guiding the public opinion is better than that of the populist or democratic leader; that is in contrast with the C-society. In general, the attraction of a leader increases in function of the leader's degree of connectivity with the network but it is not a linear effect.

With the global study of asymptotic trends in opinion evolution it is possible to detect interesting or unusual situations in opinion formation under the leaders influence. When an interesting global situation is observed, then it is just the case to study the scenery in more details, including time evolution of opinion of individuals and their positioning or connectivity in the network. Results of our modeling and simulation suggest that, besides the features mentioned above, leaders need more sophisticated strategies to maintain the control over societies characterized by a high degree of mistrust and antagonism. We think that some of the results can be applied also to the effect of mass media onto the dynamics of public opinion formation. Competition of leaders, leader's strategies, different kinds of the social power of leaders, and the network topology are expected to affect the opinion formation; those are natural extensions of this work for the future. We expect that the C/PA-model with leaders helps to explain some tendencies in opinion formation in heterogeneous societies composed of agents similar to those of the C- and PA-type.

**Acknowledgments**
Authors are grateful to Dr. Héctor Calderón Gama for helpful discussions and suggestions.

**Appendix**

Time evolution of opinions is occurred in a variety of ways, depending on the C/PA-society composition, leader's opinion $x_l$ and tolerance $u_l$, leaders connectivity $d_l$ and the average opinion uncertainty of agents $U$. So many parameters suggest the topic of the time convergence of opinions to be studied mathematically rigorously or by means of practical computer runs for particular sceneries described by a specific set of values of model parameters. Since the aim of this work is to see the global statistical trends in opinion formation in societies composed of agents of different "psychological" type, we did not use any formal mathematical criterion to control the time convergence of opinions. Based on the previous experience and some probe runs, we chose for this work an intermediate way to "control" the convergence. In each experiment we have fixed the values of the following parameters: C/PA-society composition, leader's opinion and tolerance $x_l$ and $u_l$, leaders connectivity $d_l$ and the average opinion uncertainty of agents $U$; besides, we generate new network (of the same type and same number of agents), new uniform distribution of opinions and uncertainties of agents. We let the system evolve for 500 "time" steps. Each time step involves N pair interactions. Then final opinion distributions of 50 experiments are averaged to give the trend of opinion formation (one point in a plot of Figures 1 to 7). We verified visually the convergence of the opinion formation in many sampled (not in all) runs, and observed that 500 "time" steps is reasonable evolution time to see the asymptotic opinion distribution. Moreover, if the experiments were not stable asymptotically or if there was no convergence in opinion formation, then the opinion distribution averaged over 50 experiments would give almost uniform distribution, and that is not the case.

Three figures illustrate some ways of convergence, but we observed many more interesting details in other sceneries that worth to be studied and commented rigorously in a separate work.

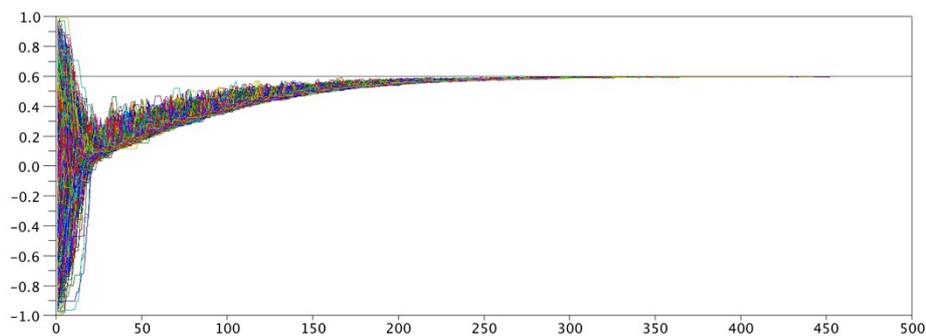

Figure A. Time evolution of opinions in a pure C-society: PA000/C100, $x_l = 0.6$, $u_l = 0.2$, $d_l = 0.6$ and $U = 0.6$.

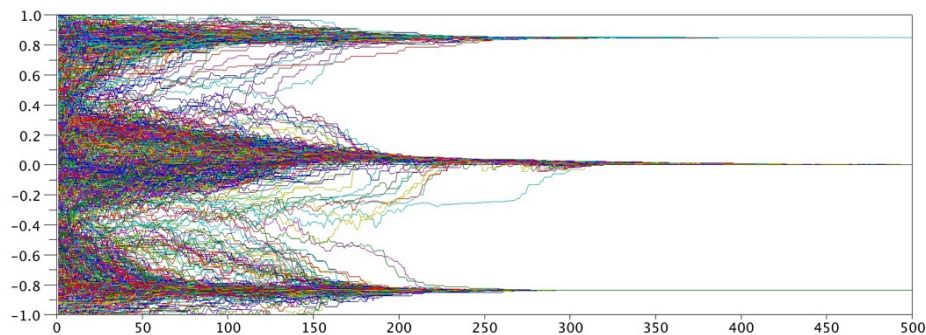

Figure B. Time evolution of opinions in a pure PA-society: PA100/C000, $x_l = 0$, $u_l = 0.5$, $d_l = 0.6$ and $U = 0.4$.

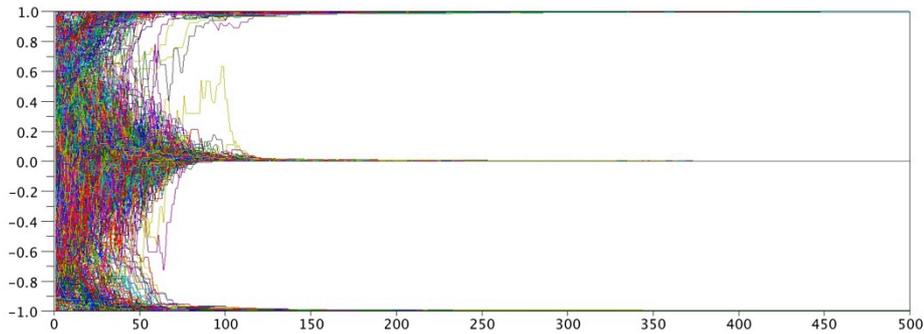

Figure C. Time evolution of opinions in a pure mixed C/PA-society: PA60/C40, $x_l = 0$, $u_l = 0.2$, $d_l = 0.2$ and $U = 0.6$.

Group opinion convergence in time was observed in all cases of sampled evolution plots, even though interesting details are observed at early stages of opinion evolution, especially. Time convergence of opinions in presence of a populist ($u_l = U$) and democratic leader ($u_l = .50$) is similar. In any case, 500 "time" steps show the trends of opinion formation clearly.

Some of the observed features of opinion evolution are as follows. **C-society** (Figure A)**:** Two stages of group opinion evolution are clearly observed. First stage is a relatively rapid self-organization of opinions in a compact cluster (20 to 50 "time" steps, depends on dl and U), and then the cluster of opinions tends to the opinion of the leader, if it is the case. **PA-society** (Figure B)**:** Opinion groups are usually formed at 50 to 120 "time" steps. **Mixed PA60/C40-society** (Figure C)**:** Self-organization of opinions of agents and adhesion of opinions of other agents to the opinion of leader are observed simultaneously. The larger $U$, the less is the convergence time. The larger $d_l$, the slower is the convergence in self-organized groups of opinion that do not share the opinion of a leader. Opinion convergence in a mixed C/PA-society is usually slower than in the case of pure C- or PA-societies at the same values of parameters.